\documentclass[figure]{./epl}

\title{Stochastic resonance in periodic potentials: realization in a
       dissipative optical lattice}   

\shorttitle{Stochastic resonance in periodic potentials}

\author{M. Schiavoni, F.-R. Carminati, L. Sanchez-Palencia,
        F. Renzoni and G. Grynberg}
\shortauthor{M. Schiavoni \etal}

\institute{Laboratoire Kastler Brossel, D\'epartement de Physique de l'Ecole
Normale Sup\'erieure, 24, rue Lhomond, 75231, Paris Cedex 05, France}

\begin{document}

\maketitle

\begin{abstract}
We have observed the phenomenon of stochastic resonance on the 
Brillouin propagation modes of a dissipative optical lattice. 
Such a mode has been excited by applying a moving potential
modulation with phase velocity equal to the velocity of the mode.  
Its amplitude has been characterized by the center-of-mass (CM) 
velocity of the atomic cloud. At Brillouin resonance, we studied 
the CM-velocity as a function of the optical pumping rate at a given 
depth of the potential wells. We have observed a resonant dependence of 
the CM velocity on the optical pumping rate, corresponding to the noise 
strength. This corresponds to the experimental observation of stochastic 
resonance in a periodic potential in the low-damping regime.
\end{abstract}

A particle trapped in a potential well constitutes a model useful for the
understanding of a variety of phenomena. The extension to a periodically 
modulated double well potential including a stochastic force leads to a 
complex nonlinear dynamics, and allows to modelize a variety of phenomena
ranging from geophysics \cite{benzi,nicolis} to bistable ring lasers 
\cite{ring}, from neuronal systems \cite{douglass} to the dithering effect 
in electronics \cite {dithering} and so on. Indeed such a system exhibits 
the phenomenon of {\it stochastic resonance} (SR \cite{moss,gamma}): the 
response of the system to the input signal (the modulation) shows a resonant 
dependence on the noise level (the amplitude of the stochastic force), so 
that an increase of the noise strength may lead to a better synchronization 
between the particle motion and the potential modulation. 

The phenomenon of stochastic resonance is not restricted to static
double-well potentials driven by a periodic and a stochastic force,
and new types of stochastic resonance have been demonstrated in various
systems, as systems with a single potential well, bistable systems 
with periodically modulated noise, and many others 
\cite{dykman,hu,fronzoni,marchesoni,kim,dan,bao}.
In particular much attention has been devoted to the analysis of SR
in {\it periodic potentials} \cite{hu,fronzoni,marchesoni,kim,dan,bao}. 
Indeed, many different physical systems are described in terms of periodic
structures, and it is by now well established that the noise plays a 
major role in the mechanisms of transport in periodic structures.
For example, the study of the underdamped motion of a particle in a
periodic potential showed that it is the interplay between inertial and
thermal effects which determines the peculiar mechanical properties of
certain metals \cite{isaac,kolomeisky}.
This is precisely the regime examined in this work: we study the 
SR phenomenon by taking as spatially periodic system a dissipative 
optical lattice \cite{robi}. The laser fields create the periodic 
potential and produce the stochastic process of optical pumping. 
The friction for atoms well localized in a potential well is very 
small, so that inertial effects are important (low-damping regime).
We report the experimental observation of stochastic resonance on the
propagation modes of a dissipative optical lattice and give a complete 
theoretical account of the experimental findings.

\begin{figure}[ht]
\begin{center}
\onefigure[width=6cm]{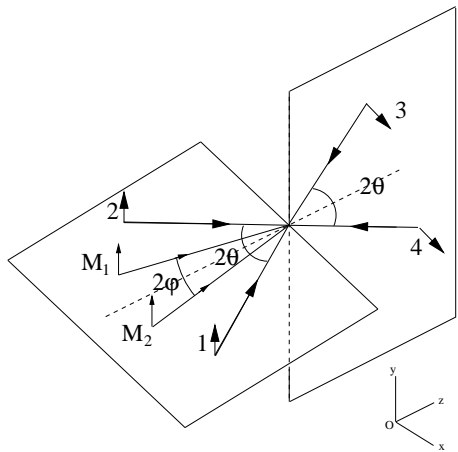}
\end{center}
\caption{Laser fields configuration for the 3D lin$\perp$lin optical lattice.
The beams $1-4$ generate the static 3D periodic potential. Two additional
laser beams (M$_1$ and M$_2$), are introduced to create a moving potential
modulation.}
\label{fig1}
\end{figure}

The three dimensional  periodic structure is generated by the interference 
of four linearly polarized laser beams, arranged in the so-called 
lin$\perp$lin configuration (Fig. \ref{fig1}) \cite{robi}. The resulting
optical potential has minima located on a orthorombic lattice and associated
with pure circular (alternatively $\sigma^{+}$ and $\sigma^{-}$) polarization.
The lattice constants, i.e. the distance (along a major axis) between two
sites of equal circular polarization are $\lambda_{x,y}=\lambda/\sin\theta$
and $\lambda_z =\lambda/(2\cos\theta)$, with $\lambda$ the laser field
wavelength, and $2\theta$ the angle between two copropagating lattice beams.

\begin{figure}[ht]
\begin{center}
\onefigure{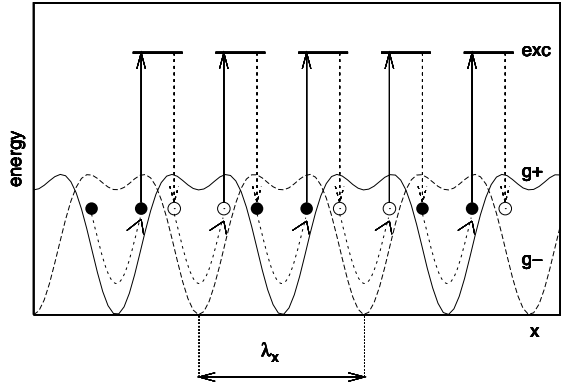}
\end{center}
\caption{Atomic trajectory corresponding to a Brillouin mode in the
$x$-direction.
The shown potential curves ($g_{+}$ and $g_{-}$) are the section along
$y=z=0$ of the optical potential for a $J_g=1/2\to J_e=3/2$ atomic transition
and a 3D lin$\perp$lin beam configuration.}
\label{fig2}
\end{figure}

The Brillouin-like propagation modes in such optical lattices have been first
identified in Ref. \cite{brillo} via semiclassical Monte Carlo simulations 
\cite{mc}. They consist of a sequence in which one half oscillation in a 
potential well is followed by an optical pumping process to a neighbouring 
well, and so on (Fig. \ref{fig2}). 
The velocity of the Brillouin mode is easily calculated by neglecting the
corrections due to the anharmonicity of the optical potential. The time
for an atom to do half an oscillation is then $\tau=\pi/\Omega_x$, where
$\Omega_x$ is the $x$-vibrational frequency.  This corresponds to an average 
velocity
\begin{equation}
\bar{v}=\frac{\lambda_x/2}{\tau}=\frac{\lambda \Omega_x}{2\pi\sin\theta}~.
\end{equation}

The direct observation of the Brillouin modes in optical lattices has been
recently reported \cite{brillo2}. We note, however, that the detection scheme
used in that work was based on the measurement of diffusion coefficients.
These measurements require averaging of a large data set, and this makes
difficult the exploration of a large interval of interaction parameters, 
as necessary to evidence the phenomenon of stochastic resonance. The
excitation scheme introduced in this work will results instead in
significant variations of the atomic cloud center-of-mass motion, and leads
to the observation of stochastic resonance, as described now.

The transport of atoms in optical lattices has been extensively studied 
\cite{leiden1,leiden2,epjd1,epjd2}. In a {\it dissipative} optical lattice
the dominant transport process is spatial diffusion \cite{epjd1,epjd2},
and the Brillouin modes are greatly suppressed. To excite these modes it is
necessary to create a potential modulation moving with phase velocity equal
to the velocity of the Brillouin mode. This is done by introducing two 
additional $y$-polarized laser fields (M$_1$ and M$_2$, see Fig. \ref{fig1}).
They propagate in the $xOz$ plane, symmetrically displaced with respect to 
the $z$-axis, and they form an angle equal to $2\varphi$. These two modulation
beams are taken to be sufficiently detuned from the lattice fields to neglect 
the interference between them and the lattice beams on the time scale of
the atomic motion. In this way the modulation interference pattern is due 
only to the two fields M$_1$ and M$_2$, and consists of an intensity
modulation moving along the $x$ axis with phase velocity 
\begin{equation}
v_{\phi}=\frac{\delta_m}{|\Delta \vec{k}|}=\frac{\delta_m}{2k_m\sin\varphi}
\end{equation}
where $\delta_m$ is the detuning between the fields M$_1$ and M$_2$, and
$\Delta \vec{k}=\vec{k}_{M_1}-\vec{k}_{M_2}$ the difference between
their wavevectors ($|\vec{k}_{M_j}|\simeq k=2\pi/\lambda$,  $j=1,2$).
This results in a moving modulation of the optical potential. For a 
$1/2\to 3/2$ atomic transition\footnote{It is customary in the analysis of 
Sisyphus cooling to consider a $1/2\to 3/2$ atomic transition \cite{robi}.},
the modulated potential for the two ground states $|\pm 1/2\rangle$ reads
\begin{equation}
U_{\pm}(\vec{r}) = U^0_{\pm}(\vec{r})+
\delta U\cdot \cos{ [(\Delta k_x x - \delta_m\cdot t)]}
\end{equation}
with $U^0_{\pm}$ the optical potential of the unperturbed lattice
\begin{equation}
U^0_{\pm}(\vec{r}) = \frac{8\hbar\Delta'_0}{3}\left[
\cos^2(k_x x) + \cos^2(k_y y) 
\mp \cos(k_x x)\cos(k_y y)\cos(k_z z) \right]~,
\end{equation}
and $\delta U=4\hbar\Delta'_{0,m}/3$
the amplitude of the potential modulation. $\Delta'_0$ ($\Delta'_{0,m}$)
denotes the light shift per lattice (modulation) field. We expect that
for $v_{\phi}=\bar{v}$, i.e. for $\delta_m = \pm \Omega_B$, with
\begin{equation}
\Omega_B \equiv \frac{2\sin\varphi}{\sin\theta}\Omega_x ~,
\label{eq:omegab}
\end{equation}
the Brillouin mode is excited, with the atoms following the potential
modulation. This has been confirmed by Monte Carlo simulations. For a 
given modulated optical potential $U_{\pm}$ and a given optical pumping 
rate $\Gamma'_0$, we calculated the position of the center of mass (CM)
of the atomic cloud as a function of the interaction time, for different
values of the detuning $\delta_m$ between the two driving fields.

\begin{figure}[ht]
\begin{center}
\onefigure{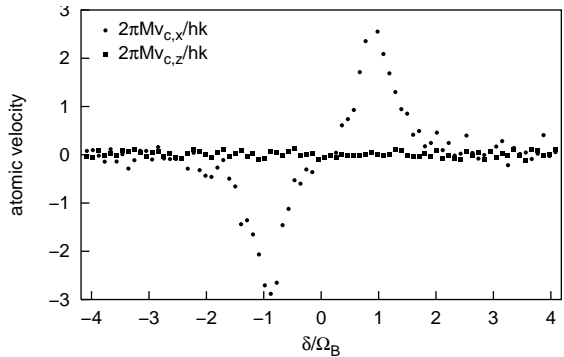}
\end{center}
\caption{Numerical results for the velocity of the CM of the atomic cloud
as a function of the detuning $\delta_m$ between the two driving fields.
The velocity is in units of recoil atomic velocity $v_r$. The lattice beam
angle is $\theta = 30^0$, the lattice detuning from atomic resonance
$\Delta=-10\Gamma$ and the light shift per beam $\Delta_0^{'}=-200\omega_r$.
Here $\Gamma$ and $\omega_r$ are the width of the excited state and the
atomic recoil frequency, respectively. The driving field angle is
$\varphi=10^0$, the detuning $\Delta_m = -30 \Gamma$ and light shift per beam
$\Delta_{0,m}'=-20\omega_r$.}
\label{fig3}
\end{figure}

The application of the moving modulation produces a motion of the CM of 
the atomic cloud. Its velocity $v_c$ strongly depends on the velocity
of the moving modulation, i.e. on the detuning $\delta_m$ between the 
driving fields, and shows two resonances centered at $\delta_m=\pm\Omega_B$ 
(Fig. \ref{fig3}). These resonances correspond to the excitation of 
the propagation mode in the $\pm x$ direction: at $\delta_m=\pm\Omega_B$ 
the velocity of the moving modulation is equal to the velocity of the 
Brillouin mode, and the atoms follow the potential modulation. On the
contrary, for a velocity of the moving modulation very different from 
the velocity of the Brillouin mode ($|\delta_m |\gg\Omega_B$ or 
$|\delta_m |\ll\Omega_B$) the atomic dynamics is left unperturbed, and the 
CM of the atomic cloud does not move. This analysis shows that the 
effective excitation of the Brillouin propagation modes can be detected 
by observing a displacement of the CM of the atomic cloud. 
This will be the strategy followed in our experiment. We verified that 
the excitation of the Brillouin modes also leads to a resonant increase 
of the diffusion coefficient in the $x$-direction, in agreement with 
previous results for a different modulation scheme \cite{brillo2}.

In our experiment, $^{87}$Rb atoms are cooled and trapped in a magneto
optical trap. The trapping beams and the magnetic field are then suddenly
turned off. Simultaneously the four lattice beams are turned on. After
10 ms of thermalization of the atoms in the lattice the two laser fields
for the moving modulation are introduced according to the geometry of
Fig. \ref{fig1}. The lattice angle is $\theta=30^0$, while the two driving
fields form an angle $2\varphi=37^0$. The two driving fields are derived
from an additional laser, with their relative detuning controlled by
acousto-optical modulators. 

\begin{figure}[ht]
\begin{center}
\onefigure{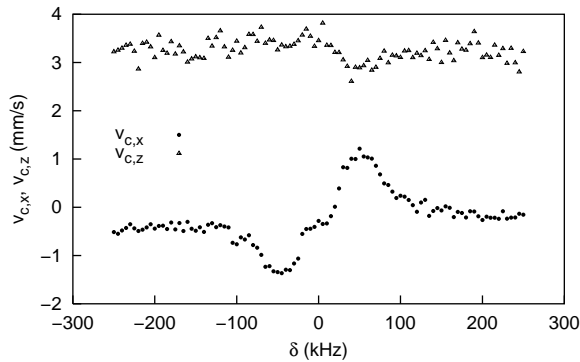}
\end{center}
\caption{Experimental results for the velocity of the CM of the
atomic cloud as a function of the detuning $\delta_m$ between
driving fields. The lattice parameters are: lattice detuning
$\Delta/(2\pi) = -45.6$ MHz, intensity per lattice beam $I=2.3$ mW/cm$^2$,
lattice angle $\theta=30^0$. These parameters correspond to a vibrational
frequency in the $x$- direction $\Omega_x/(2\pi) \simeq 45$ kHz.
The parameters for the moving modulation are:
$I_{M1} \simeq I_{M2} \simeq 0.5$ mW/cm$^2$, $\Delta_m/(2\pi) = -44$ MHz,
$2\varphi=37^o$.  From these data we derive through Eq. (\ref{eq:omegab})
$\Omega_B \simeq 2\pi\cdot 55$ KHz, in excellent agreement with the
experimental findings.}
\label{fig:exp:brillo}
\end{figure}

The transport of the atoms in the optical lattice is studied by direct
imaging of the atomic cloud with a CCD camera. We verified that for a
given detuning $\delta_m$, i.e. for a given velocity of the moving potential
modulation, the motion of the center of mass of the atomic cloud is 
uniform and correspondingly determined the CM velocity $v_c$. 
By repeating the measurements for different detunings between driving
fields, we obtained the $x$- and $z$-component of the CM-velocity $v_c$ 
as a function of $\delta_m$, as reported in Fig. \ref{fig:exp:brillo}.
The $x$-component shows a resonant behaviour with the detuning $\delta_m$, 
with two resonances of opposite sign symmetrically displaced with respect
to $\delta_m=0$. The position of these resonances is in agreement with the
value $\Omega_B\simeq 2\pi\cdot 55$ kHz derived from the lattice parameters
via Eq.  (\ref{eq:omegab}). In contrast, the data for the $z$-component 
$v_{c,z}$, whose offset value corresponds to the radiation pressure of the 
modulation fields, do not show any resonance. These results are in agreement 
with our numerical simulations and constitute the direct experimental 
observation of the Brillouin propagation modes via the detection of the 
displacement of the CM of the atomic cloud.

\begin{figure}[ht]
\begin{center}
\onefigure{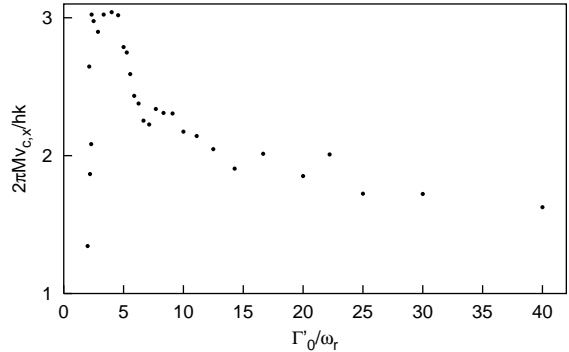}
\end{center}
\caption{Numerical results for the $x$-component of the velocity of the
center of mass of the atomic cloud as a function of the optical pumping
rate, for a given depth of the optical potential wells. Parameters of
the calculations are:
$\theta=30^0$ and $\Delta_0^{'}=-100\omega_r$, $\varphi=10^0$,
$\Delta_m = -30\Gamma$ and $\Delta^{'}_{0,m} = -10\omega_r$.}
\label{fig:th:stoc}
\end{figure}

The Brillouin propagation modes are determined by the synchronization of
the oscillations within a potential well with the hopping from a well to 
a neighbouring one produced by the optical pumping \footnote{The propagation
mechanism associated with these modes differs from that encountered in dense 
fluids or solid media. The atomic density is so low that the interaction 
between the different atoms is completely negligible, therefore the mechanism
for the propagation of atoms cannot be ascribed to any sound-wave-like 
mechanism.}\footnote{A non-zero current in a symmetric periodic
potential can also be obtained by modifying, through an external driving 
field, the activation energies of escape from a well, as described in
\cite{dd1,dd2}. However that mechanism of directed diffusion does not
correspond to the propagation of atoms at a well defined velocity, as in
our case.}. We studied the amplitude of the Brillouin mode, here 
characterized by the velocity of the CM of the atomic cloud 
$v_c(\delta_m=+\Omega_B)$ [analogous results are obtained for 
$v_c(\delta_m=-\Omega_B)$], as a function of the optical pumping rate 
$\Gamma_0^{'}$ for a given modulated optical potential. The numerical results 
display the SR-like nonmonotonic dependence of the amplitude of the Brillouin 
mode on the noise strength (Fig. \ref{fig:th:stoc}), in agreement with our
previous results for a different modulation scheme \cite{brillo2}. This 
SR-scenario has one important peculiarity with respect to the
model usually considered in the analysis of stochastic resonance.
Stochastic resonance is in general understood as the noise-induced
enhancement of a weak periodic signal with a frequency much smaller than
the intrawell relaxation frequency within a single metastable state. In
contrast, in the present case, the noise synchronizes precisely with the
intrawell motion of the atoms.

\begin{figure}[ht]
\begin{center}
\onefigure{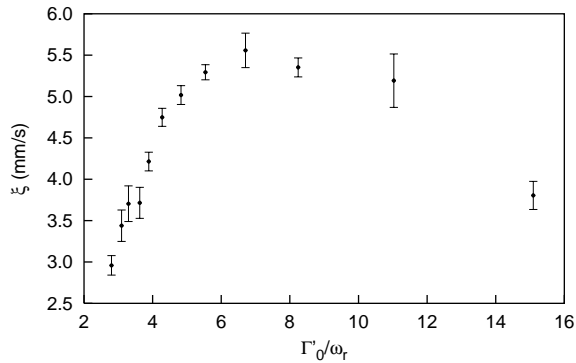}
\end{center}
\caption{Experimental results for the peak-to-peak amplitude $\xi$
of the CM velocity curve, as a function of the optical pumping rate
$\Gamma_0'$, at a given depth of the potential wells and given
amplitude modulation. The light shift per lattice beam is
$\Delta'_0=-37.5\omega_r$. The parameters of the laser fields creating
the moving intensity modulation are the same as for Fig.
\protect\ref{fig:exp:brillo}.}
\label{fig6}
\end{figure}

Although $v_{c,x}(\delta_m=+\Omega_B)$ and $v_{c,x}(\delta_m=-\Omega_B)$ are 
expected to have the same dependence on the optical pumping rate, 
experimentally it is more convenient to characterize the amplitude of 
the propagation mode by the peak-to-peak amplitude $\xi$
\begin{equation}
\xi = v_{c,x} (\delta_m=+\Omega_B) - v_{c,x} (\delta_m=-\Omega_B)
\label{eq:xi}
\end{equation}
of the CM velocity curve (as the one of Fig. \ref{fig:exp:brillo}).
By doing so, the eventual uniform drift of the atomic cloud along the $x$-
direction as a result of the radiation pressure deriving from a small
difference in the driving fields intensities does not affect our
measurements.  We studied the $\xi$ parameter at Brillouin resonance as 
a function of the optical pumping rate $\Gamma'_0$ at a given depth 
of the potential wells. This has been done by varying the lattice intensity 
$I$ and detuning $\Delta$ so to keep the depth of the potential wells 
$U_0\propto I/\Delta$ constant while varying the optical pumping rate 
$\Gamma'_0 \propto I/\Delta^2$. The intensity and the detuning 
$\Delta_m$ of the modulation fields are instead kept constant.
Results of our measurements of $\xi$ as a function of the optical pumping
rate at a given depth of the potential wells and given modulation are shown
in Fig. \ref{fig6}.

The typical behaviour of SR is observed: the parameter
$\xi$ increases with $\Gamma'_0$ at low pumping rates; then a maximum
is reached corresponding to the synchronization between the optical
pumping from one well to the next one with the oscillation in the potential
wells; finally at larger pumping rates this synchronization is lost and
$\xi$ decreases.

In conclusion, we reported the observation of stochastic resonance 
on the Brillouin modes of a dissipative optical lattice. These modes 
have been excited by applying a moving potential modulation with 
phase velocity $v_{\phi}$ equal to the velocity $\bar{v}$ of the 
Brillouin mode. This results in a motion of the center of mass of
the atomic cloud. The effective excitation of the Brillouin propagation
mode has been detected by observing a resonant dependence of the 
velocity of the atomic cloud CM on the velocity of the moving modulation,
with a maximum CM-velocity at $v_{\phi}=\bar{v}$. To observe the 
phenomenon of stochastic resonance in the optical lattice, we studied 
the CM-velocity at Brillouin resonance as a function of the optical
pumping rate at a given depth of the potential wells and a given
modulation amplitude.  The SR-like nonmonotonic dependence of the
CM-velocity on the optical pumping rate has been observed. 

We thank Yanko Todorov for useful comments on the manuscript.
This work was supported by the CNRS and the R\'egion Ile de France 
under contract E.1220 "Atomes ultrafroids: vers de nouveaux \'etats 
de la mati\`ere". 
Laboratoire Kastler Brossel is an "unit\'e mixte de recherche de
l'Ecole Normale Sup\'erieure et de l'Universit\'e Pierre et Marie
Curie associ\'ee au Centre National de la Recherche Scientifique
(CNRS)".

\end{document}